\documentclass[a4paper,fleqn]{cas-sc}

\def\tsc#1{\csdef{#1}{\textsc{\lowercase{#1}}\xspace}}
\tsc{WGM}
\tsc{QE}
\tsc{EP}
\tsc{PMS}
\tsc{BEC}
\tsc{DE}

\begin{document}
\let\WriteBookmarks\relax
\def\floatpagepagefraction{1}
\def\textpagefraction{.001}
\shorttitle{Saturated absorption and electromagnetically induced transparency} 
\shortauthors{A. Sargsyan, et.al.}

\title [mode = title]{Saturated absorption and electromagnetically induced transparency of residual rubidium in dense cesium vapor}




\author[1]{Armen D. Sargsyan}[orcid=0000-0003-2138-7462]

\credit{Investigation, Writing -- review \& editing, Funding acquisition} 

\affiliation[1]{organization={Institute for Physical Research, National Academy of Sciences of Armenia},
                addressline={Gitavan-2}, 
                city={Ashtarak},
                postcode={0204}, 
                country={Armenia}}

\author[1]{Anahit L. Gogyan}[orcid=0000-0002-5629-9959]
\ead{agogyan@gmail.com}
\cormark[1]
\credit{Software, Formal analysis} 

\author[1]{David H. Sarkisyan}[orcid = 0000-0002-3432-0796]

\credit{Writing -- review \& editing, Writing -- original draft, Supervision, Funding acquisition} 

\cortext[cor1]{Corresponding author}

\begin{abstract}
In the sealed-off cesium vapor cell studied in this work, a residual rubidium fraction of approximately $\sim$1\% was observed. We investigate the optical response of these trace Rb atoms in a sealed 1~cm long Cs-filled vapor cell. Despite the low concentration, laser excitation at 795~nm allows the observation of saturated absorption and electromagnetically induced transparency (EIT) resonances. The surrounding Cs vapor effectively acts as a buffer medium, reducing the Rb atomic velocity and increasing the interaction time with the laser field, which improves the EIT signal. The experiments are performed in an all-sapphire cell that can be heated up to 500$^{\circ}$C without window blackening, unlike conventional glass cells. From the measured spectra, Cs--Rb collisional cross sections are estimated. These results show that residual atomic species in high-temperature vapor cells can be exploited for spectroscopic and nonlinear-optical studies.
\end{abstract}


\maketitle

\section{Introduction}

Atoms of alkali metals (Cs, Rb, etc.) possess strong optical transitions in the visible and near-infrared spectral range (600--900~nm), where tunable lasers with excellent spectral properties are readily available. This fact underlies the wide use of vapor cells with thicknesses ranging from hundreds of nanometers to several centimeters, filled with alkali atomic vapors, in many areas of physics including precision metrology, magnetometry, and optical telecommunications~\cite{1,2,3,4,5,6p}. More recently, alkali atoms have also been extensively employed in studies of highly excited Rydberg states, which exhibit extreme sensitivity to external electric and magnetic fields~\cite{6}. 

One of the principal requirements for optical vapor cells is resistance to chemically aggressive hot alkali vapors, particularly when high atomic densities are required. Earlier studies reported that sealed Rb and Cs cells may contain trace residual fractions of the other alkali species (on the order of 0.01\%)~\cite{7,8}. However, high-resolution spectroscopy of such residual atoms has not been realized so far.

In the present work, we employ a centimeter-long T-shaped optical cell made entirely of technical sapphire (Al$_2$O$_3$), referred to as an all-sapphire cell (ASC), filled with Cs atomic vapor. Such cells can be heated up to 500$^{\circ}$C without surface degradation or window blackening~\cite{9}. Operation at elevated temperature is essential to reach sufficient vapor density for detecting optical signals from a small fraction of residual metal atoms (or a small abundance of a specific isotope). In the sealed ASC used here, the presence of approximately 1\% Rb (with Cs as the dominant component) was detected. We show that even this small fraction is sufficient to enable high-resolution spectroscopy of Rb atoms.

For the first time, saturated absorption (SA) spectroscopy of residual rubidium vapor (RRV) has been performed in Cs-filled cells of thickness $L=1$~cm and $L=40~\mu$m, and the results obtained in centimeter- and micrometer-scale cells are compared. In addition, the coherent process of electromagnetically induced transparency (EIT) in the $\Lambda$ system of Rb atoms is investigated, and a comparison between coherent and non-coherent processes is presented. In this configuration, the surrounding Cs vapor effectively acts as a buffer medium for Rb atoms, reducing their velocity and increasing the interaction time with the laser beam, which favors the formation of narrow EIT resonances. Possible applications of this approach are also discussed.

\section{Experimental arrangement}

\begin{figure}
	\centering
	\includegraphics[width=.4\textwidth]{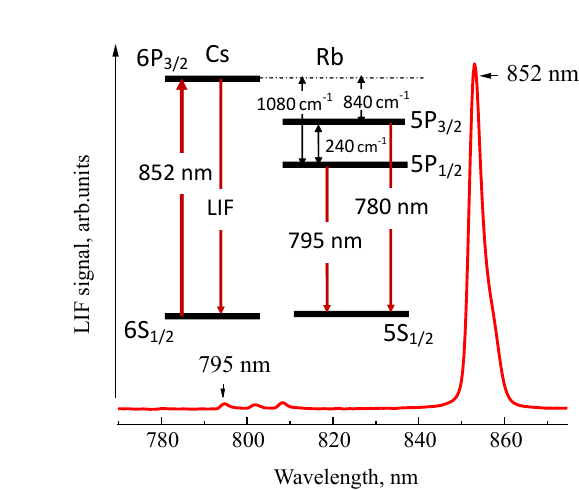}
	\caption{Laser-induced fluorescence spectrum recorded in the Cs-filled ASC under excitation at 852~nm. The dominant peak corresponds to the Cs $6P_{3/2}\rightarrow6S_{1/2}$ transition, while the weaker line at 795~nm originates from residual Rb atoms ($5P_{1/2}\rightarrow5S_{1/2}$). Inset: relevant Cs and Rb energy levels and their separation.}
	\label{FIG:1}
\end{figure}

The experiment consists of three stages: (i) identification of residual rubidium vapor (RRV) in Cs atomic vapor, (ii) implementation of saturated absorption (SA) spectroscopy of RRV using cells of thickness $L=1$~cm and $L=40~\mu$m, and (iii) investigation of coherent (EIT) and non-coherent optical processes in RRV under excitation by two resonant laser fields.

The experiments were performed using a homemade T-shaped ASC of length 1~cm containing Cs atomic vapor, described in Ref.~\cite{9}. The cell was mounted in a specially designed oven equipped with two independent heaters: one for the cell body and one for the side-arm reservoir containing the Cs and residual Rb metal source. Owing to this design, the Cs vapor pressure is determined by the temperature at the boundary of the metal column in the side-arm reservoir.
\begin{figure}
	\centering
	\includegraphics[width=.5\textwidth]{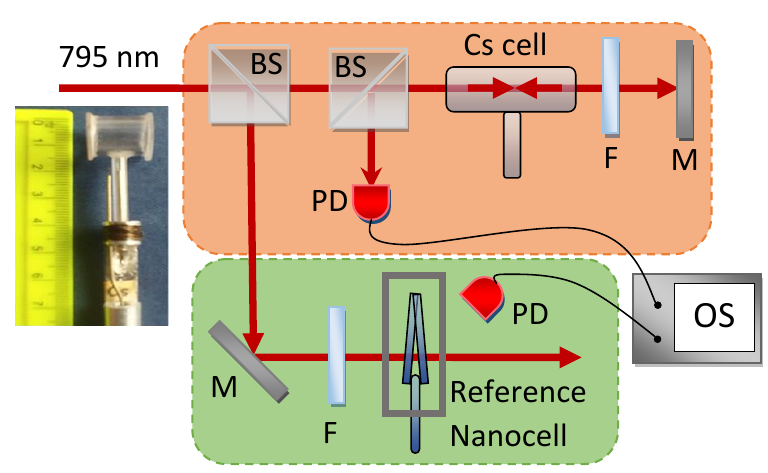}
	\caption{Experimental arrangement used for spectroscopy at 795~nm. An external-cavity diode laser (ECDL) provides the radiation interacting with a $L=1$~cm all-sapphire cell containing Cs vapor with $\sim$1\% residual Rb. M: mirror used to form counter-propagating beams, F: Filters, BS: Beam splitters, OS: Oscilloscope. The fluorescence spectrum from a Rb-filled nanocell is employed as a frequency reference.}
	\label{FIG:2}
\end{figure}

The oven provides three optical access ports: two aligned with the laser beam for transmission measurements, and a third port oriented perpendicular to the beam propagation direction for detection of laser-induced fluorescence (LIF).

\section{Detection of residual rubidium in cesium vapor}

To determine the presence of residual rubidium vapor, laser-induced fluorescence spectra were recorded, allowing simultaneous observation of Cs and Rb emission from the cell. The LIF signal was collected and analyzed using an ASEQ Instruments HR1 spectrometer covering the 200--1100~nm spectral range with a resolution of 0.4~nm.

\begin{figure}
	\centering
	\includegraphics[width=.5\textwidth]{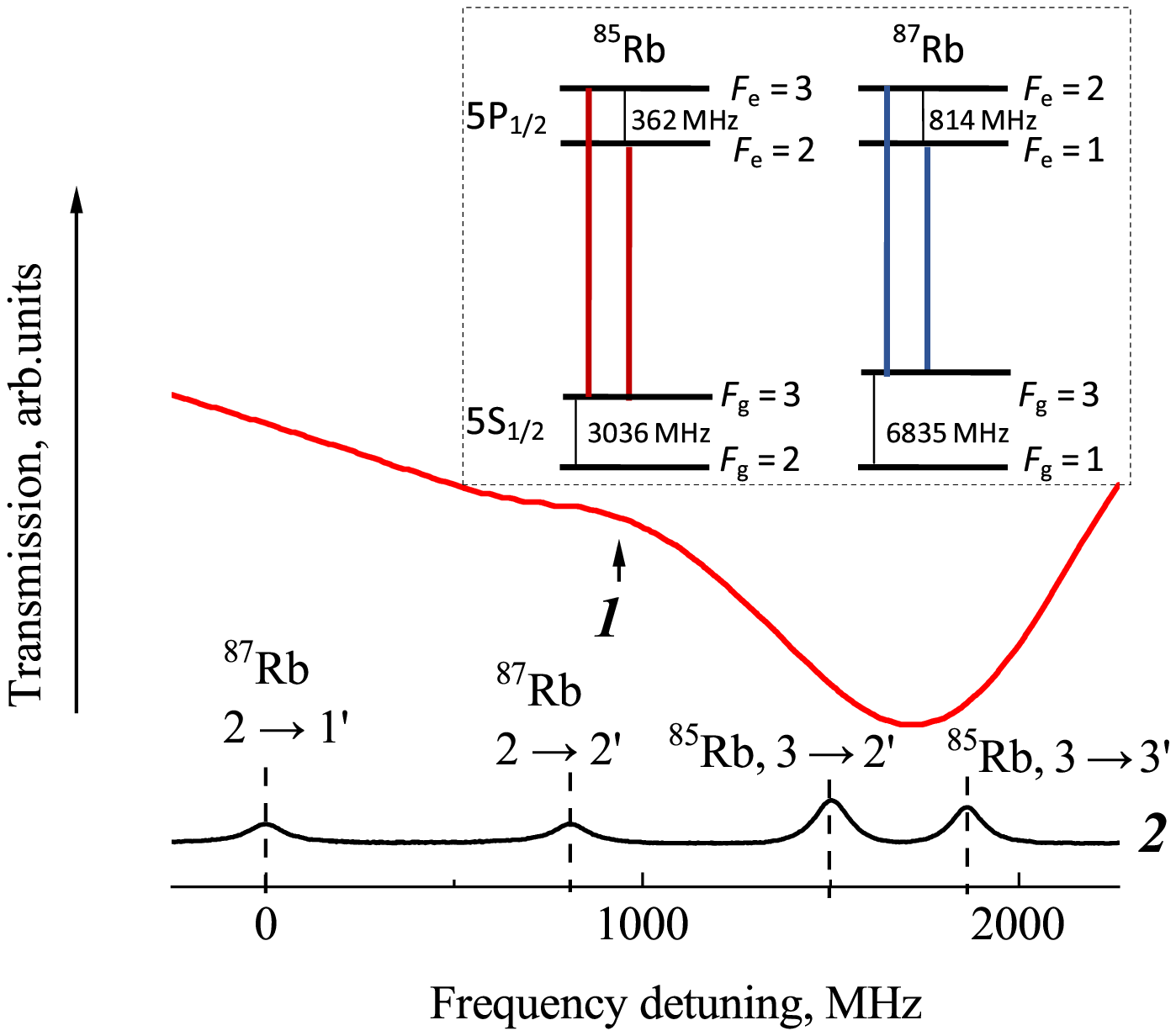}
	\caption{Curve 1 shows the Doppler-broadened transmission spectrum of 795\,nm radiation through the ASC at a temperature of 190$^\circ$C. Curve 2 presents the reference fluorescence spectrum of the Rb nanocell. The inset shows the energy-level diagram and the studied transitions of the Rb $D_1$ line.}
	\label{FIG:3}
\end{figure}
As shown previously, energy-pooling collisions enable excitation of alkali atoms from the ground state to higher-lying excited states, resulting in broadband fluorescence emission~\cite{10,11,12}. Figure~1 presents the recorded LIF spectrum when the Cs vapor is excited by laser radiation at 852~nm, which populates Cs $6P_{3/2}$ level. This excitation produces strong fluorescence at the $6P_{3/2}\rightarrow 6S_{1/2}$ transition (852~nm).

In addition to the intense Cs fluorescence, weak emission from Rb atoms is observed at 795~nm, corresponding to the $5P_{1/2}\rightarrow 5S_{1/2}$ transition. The ratio of fluorescence amplitudes $\mathrm{LIF}_{\mathrm{Cs}}(852\,\mathrm{nm})/\mathrm{LIF}_{\mathrm{Rb}}(795\,\mathrm{nm})$ indicates the presence of approximately 1\% Rb, with Cs constituting the dominant fraction in the sealed ASC. A weaker Rb fluorescence line at 780~nm ($5P_{3/2}\rightarrow 5S_{1/2}$) is also detected, with an intensity about seven times smaller than that of the 795~nm line. These observations confirm that energy-pooling collisions of excited Cs atoms transfer energy to Rb atoms, thereby populating the Rb excited states.
\begin{figure}
	\centering
	\includegraphics[width=.5\textwidth]{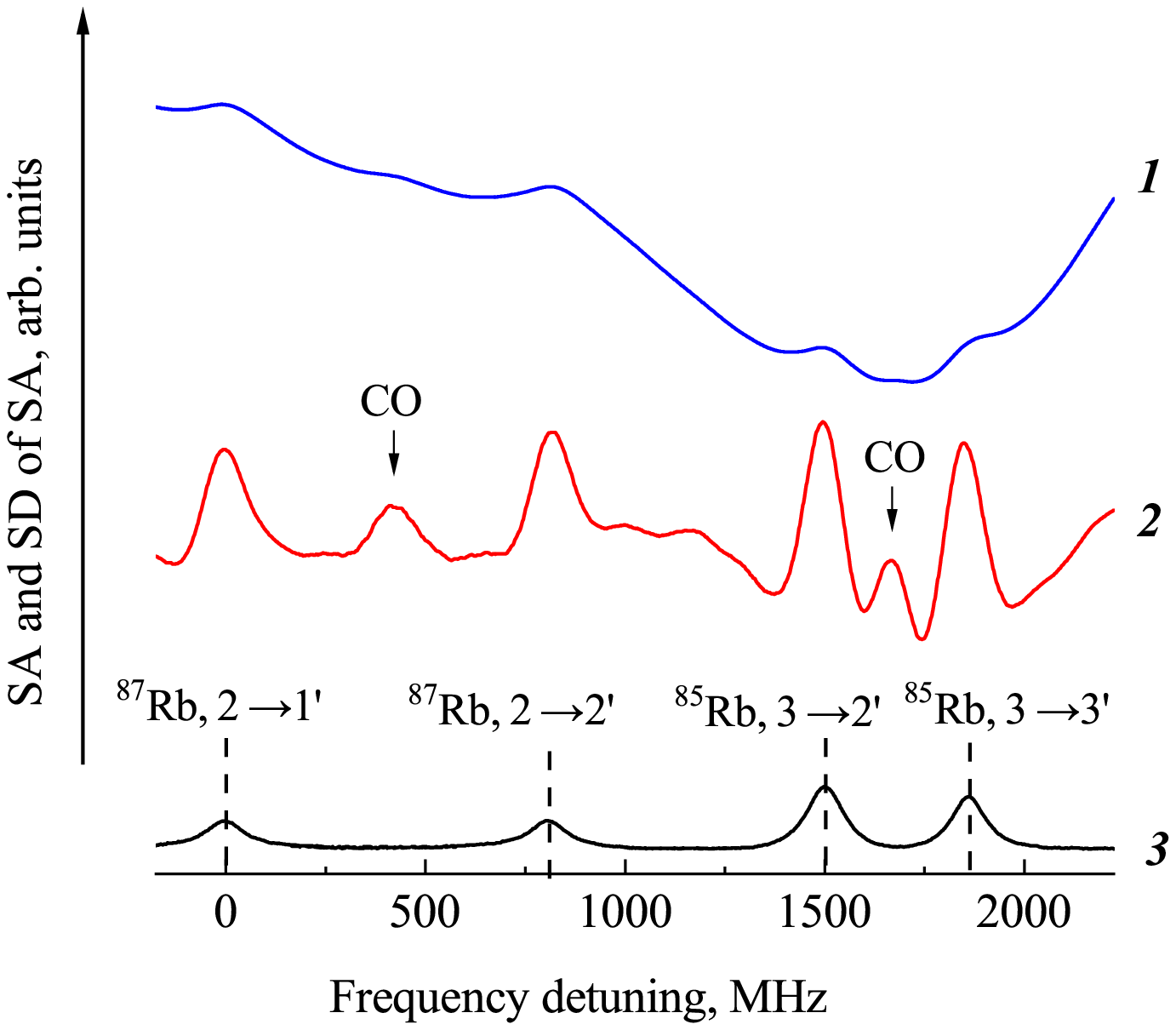}
	\caption{Curve 1 shows the experimentally recorded SA-broadened transmission spectrum of 795~nm radiation through the ASC at a temperature of 190~$^{\circ}$C. Curve 2 shows the second-derivative (SD) spectrum of Curve 1 and is inverted for clarity. Curve 3 shows the reference spectrum. CO denotes the cross-over resonances appearing in saturated absorption spectra~\cite{14,15,16}.}
	\label{FIG:4}
\end{figure}

The energy separations between the Cs $6P_{3/2}$ level and the Rb $5P_{1/2}$ and $5P_{3/2}$ levels are approximately 1080~cm$^{-1}$ and 840~cm$^{-1}$, respectively, as illustrated in the inset of Fig.~1. Population of the Rb $5P_{1/2}$ state can therefore occur via energy-pooling collisions between Cs atoms excited by the 852~nm radiation and ground-state Rb atoms~\cite{11,12}. Since the $5P_{1/2}$ level lies about 240~cm$^{-1}$ below the $5P_{3/2}$ level, it can be populated more efficiently through collisional processes.

\section{Saturated absorption spectroscopy of residual rubidium vapor}

Because the hyperfine structure of the Rb D$_1$ line is simpler than that of the D$_2$ line (the excited-state hyperfine splitting is larger), individual transitions can be resolved more easily. Therefore, the D$_1$ line of Rb atoms was chosen for the present study. Figure~2 shows the scheme used to implement saturated absorption (SA) spectroscopy in RRV. An external-cavity diode laser (ECDL) with a linewidth of $\sim$2~MHz, beam diameter of 3~mm, and tunable around $\lambda=795$~nm was employed to excite the $5S_{1/2}\rightarrow 5P_{1/2}$ transition.

\begin{figure}
	\centering
	\includegraphics[width=.5\textwidth]{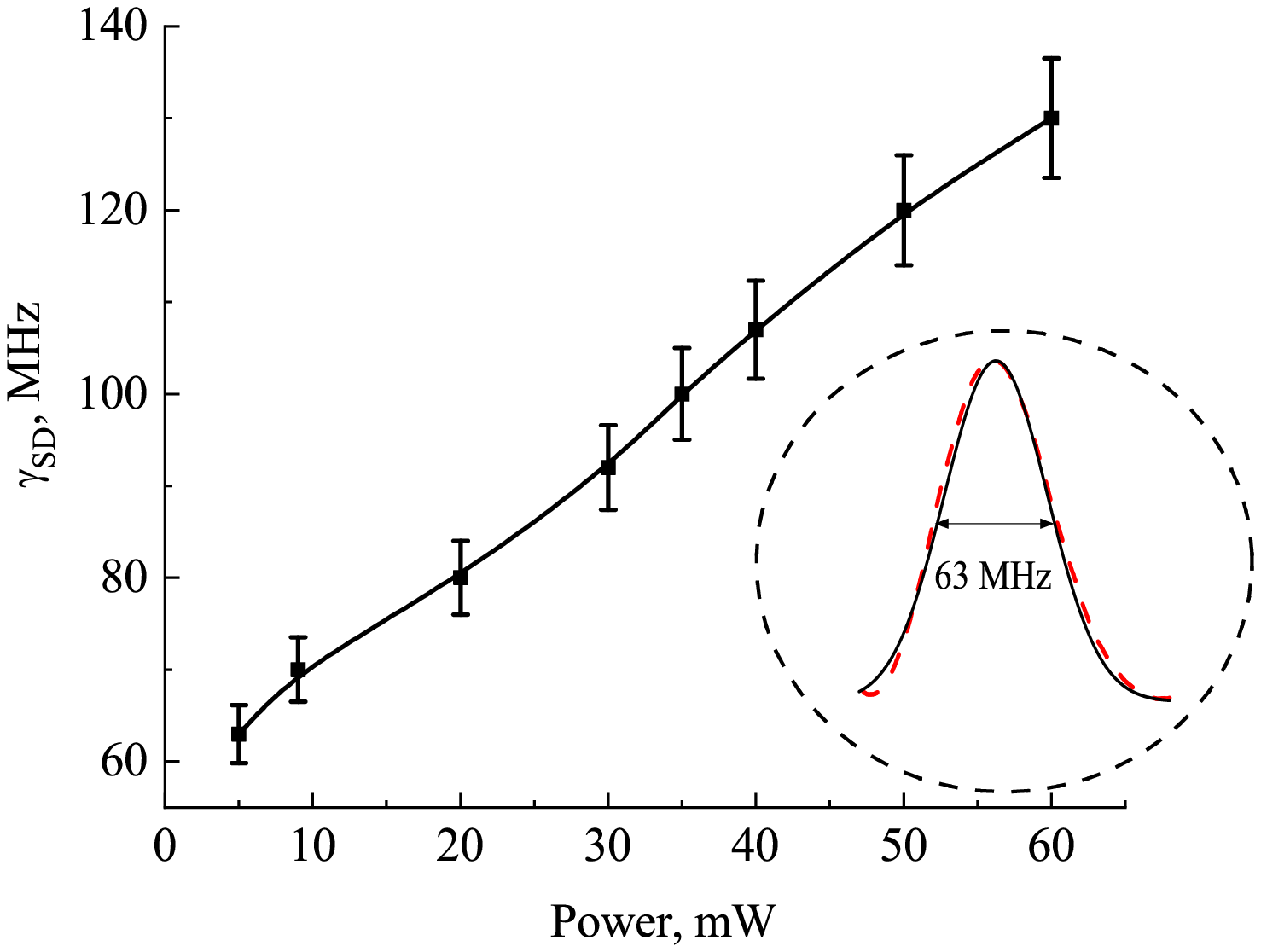}
	\caption{Dependence of the linewidth of the $^{87}$Rb $2 \rightarrow 2'$ transition on the 795~nm laser power. The inset shows a Gaussian fit of the VSOP resonance. The cell temperature is $T = 190~^{\circ}$C.}
	\label{FIG:5}
\end{figure}

Curve~1 in Fig.~3 shows the Doppler-broadened transmission spectrum of the 795~nm radiation through the ASC at a temperature of $190^{\circ}$C. The spectral profile is additionally broadened by Cs--Rb collisions. Under these conditions, the Cs atomic density is $N_{\mathrm{Cs}}\sim10^{15}$~cm$^{-3}$, while the Rb density is approximately two orders of magnitude smaller. Curve~2 in Fig.~3 presents the fluorescence spectrum recorded in a nanometric-thickness cell (NC) with a vapor column length $L=400$~nm. As shown in Ref.~\cite{13}, such NCs provide Doppler-free spectra of alkali transitions and therefore serve as convenient frequency references.

The SA scheme was implemented using counter-propagating laser beams~\cite{14,15,16}. A mirror with nearly 100\% reflectivity at 795~nm was used to generate the backward-propagating beam (Fig.~2). Curve~1 in Fig.~4 shows the experimentally measured SA spectrum in RRV at a cell temperature of $190^{\circ}$C and laser power of 40~mW.

It has been demonstrated in Refs.~\cite{17,18} that applying the second-derivative (SD) method to the recorded spectrum enhances contrast and reduces the apparent linewidth. Accordingly, after scanning the laser frequency across the resonance, the second derivative $A''(\nu)$ (curve~2 in Fig.~4) was calculated from the measured SA spectrum $A(\nu)$ (curve~1). 
As a result, the transition linewidth in the SD spectrum (see inset of Fig.~5) is reduced to $\gamma_{\mathrm{SD}}\approx60$~MHz, nearly an order of magnitude narrower than the Doppler width. This allows individual transitions to be clearly resolved. In particular, the transitions $^{87}$Rb $F=2\rightarrow F'=1,2$ and $^{85}$Rb $F=3\rightarrow F'=2,3$ are distinctly visible in curve~2 of Fig.~4. Figure~5 shows the dependence of $\gamma_{\mathrm{SD}}$ on the laser power.
\begin{figure}
	\centering
	\includegraphics[width=.5\textwidth]{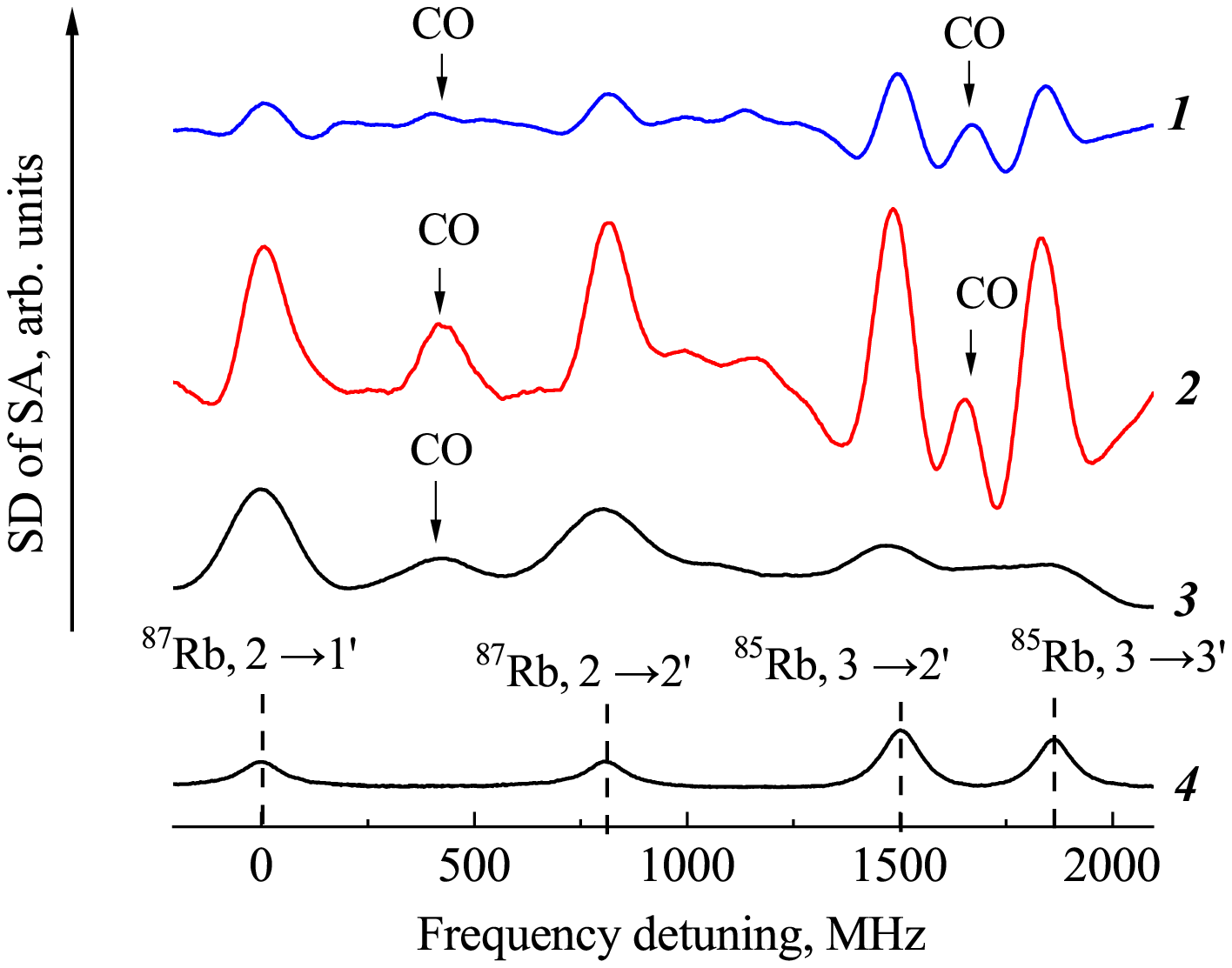}
	\caption{Saturated absorption (SA) spectra in RRV as a function of Rb atomic density. Curves 1, 2, and 3 correspond to temperatures of $150~^{\circ}$C, $190~^{\circ}$C, and $300~^{\circ}$C, respectively. Curve 4 shows the reference spectrum.}
	\label{FIG:6}
\end{figure}

Figure~6 presents SA spectra recorded at ASC temperatures of $150^{\circ}$C, $190^{\circ}$C, and $300^{\circ}$C (curves~1--3). Velocity-selective optical pumping (VSOP) resonances remain observable up to $300^{\circ}$C, corresponding to a Cs vapor pressure of about 2~Torr~\cite{15}. The lower trace in Fig.~6 shows the NC fluorescence spectrum for comparison. Figure~7 displays the dependence of $\gamma_{\mathrm{SD}}$ on the Rb density, while the inset shows its dependence on Cs density (the Rb density remains approximately two orders of magnitude lower than that of Cs over the entire temperature range).

The contribution of Rb--Rb collisions to the VSOP linewidth is $\Gamma_{\mathrm{Rb-Rb}}=KN$, where $K/2\pi=(1.10\pm0.17)\times10^{-16}$~GHz\,cm$^{3}$~\cite{19}. For the Rb density at $300^{\circ}$C this yields $\Gamma_{\mathrm{Rb-Rb}}\sim30$~MHz. The broadening due to Cs--Rb collisions can be estimated as
\begin{equation}
    \Gamma_{\mathrm{Cs-Rb}}=\sigma_{\mathrm{Cs-Rb}}\,N_{\mathrm{Cs}}\,V_{\mathrm{Cs-Rb}},
\end{equation}
where $\sigma_{\mathrm{Cs-Rb}}\sim10^{-13}$~cm$^{2}$, $N_{\mathrm{Cs}}\approx4\times10^{16}$~cm$^{-3}$, and the mean relative velocity is $V_{\mathrm{Cs-Rb}}\approx3.7\times10^{4}$~cm\,s$^{-1}$~\cite{20}. This gives $\Gamma_{\mathrm{Cs-Rb}}\sim150$~MHz, and a total collisional broadening of $\Gamma_{\mathrm{tot}}\approx180$~MHz. From these measurements and also taking into account the measurement error, we obtain a Cs--Rb collisional cross section $\sigma_{\mathrm{Cs-Rb}} \simeq (1\pm 0.1)\times 10^{-13}$~cm$^{2}$, consistent with values reported for alkali collisions with buffer gases~\cite{21}.

It is worth noting that even a small amount of buffer gas can suppress SA resonances. For example, Ref.~\cite{20} shows that adding only 0.11~Torr of Ar to a Rb cell eliminates the SA spectrum. Formation of narrow VSOP resonances requires atoms moving nearly perpendicular to the laser beams ($\mathbf{v}\perp\mathbf{k}$). One beam optically pumps this velocity group, while the counter-propagating beam detects the reduced absorption~\cite{15}. Efficient pumping requires a sufficiently long interaction time $\tau$, proportional to the transit time through the beam. Collisions with buffer-gas atoms change the atomic velocity direction, reducing $\tau$ and suppressing VSOP formation through velocity-changing collisions (VCC)~\cite{20}. 
\begin{figure}
	\centering
	\includegraphics[width=.5\textwidth]{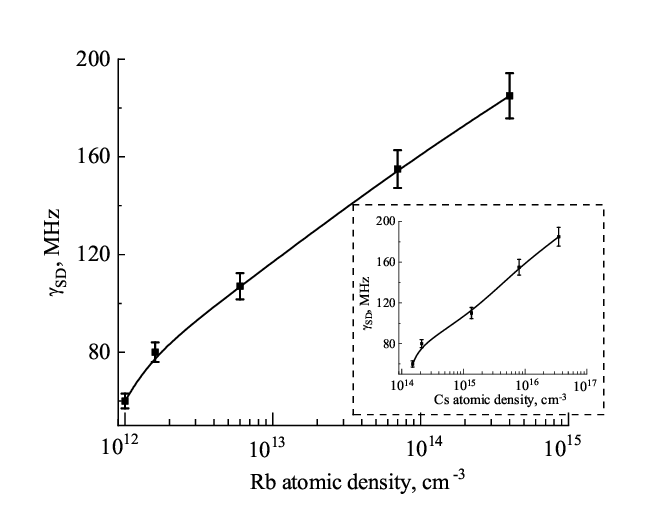}
	\caption{Dependence of the $^{87}$Rb $2 \rightarrow 2'$ transition linewidth $\gamma_{\mathrm{SD}}$ on the Rb atomic density. The inset shows the dependence of $\gamma_{\mathrm{SD}}$ on the Cs atomic density.}
	\label{FIG:7}
\end{figure}

In the present case, Cs vapor acts as the buffer medium; however, up to Cs pressures of about 2~Torr, the VCC effect remains weaker than in the Ar-buffer case, and VSOP resonances remain observable (Fig.~6).

RRV spectroscopy was also performed in a micrometer-thickness ASC cell similar to that described in Ref.~\cite{22}. Figure~8 (curve~1) shows the SA spectrum for a cell thickness of $L=40~\mu$m, while curve~2 shows its second derivative. The transitions $^{87}$Rb $F=2\rightarrow F'=1,2$ and $^{85}$Rb $F=3\rightarrow F'=2,3$, as well as crossover (CO) resonances, are clearly visible.

The ratio of the CO and VSOP amplitudes, $A(\mathrm{CO})/A(\mathrm{VSOP})$ (with $A(\mathrm{VSOP})$ corresponding to the $^{85}$Rb $F=3\rightarrow F'=2$ transition), equals 0.32 for the $L=1$~cm cell and 0.16 for the $L=40~\mu$m cell at the same temperature. As the cell thickness decreases, wall collisions increasingly suppress the CO resonances. It was shown in Ref.~\cite{23} that in ultrathin cells ($L<5~\mu$m) CO resonances disappear completely. This is explained by the fact that CO resonances originate from atoms propagating along the laser beams~\cite{15}. The transit time from the cell center to the wall is $\tau=L/V$, where $V\approx200$~m\,s$^{-1}$ is the thermal velocity, giving only a few nanoseconds; after wall collision the atom returns to the ground state, preventing CO formation. In contrast, the interaction time relevant for VSOP formation is $\tau=D/V$, where $D\sim3$~mm is the beam diameter, nearly two orders of magnitude larger, allowing VSOP resonances to persist.

\section{Electromagnetically induced transparency and optical pumping in residual rubidium vapor}

Electromagnetically induced transparency (EIT) is a quantum-interference effect in which an otherwise absorbing medium becomes transparent to a probe field in the presence of a coherent coupling field~\cite{24,25,26}. The phenomenon arises from the formation of a coherent superposition of atomic states that suppresses absorption of the probe radiation. EIT has found applications in slow-light propagation, quantum memory, and enhanced nonlinear optical processes~\cite{4,24,25,26,27,28}.

To realize EIT in residual rubidium vapor (RRV), two tunable external-cavity diode lasers (ECDLs) with linewidths of about 2~MHz and beam diameters of 3~mm were used, both tuned near $\lambda\approx795$~nm to address the Rb D$_1$ transition. The laser beams were directed approximately normal to the 1~cm long ASC containing Cs vapor. The $\Lambda$-scheme employed for $^{85}$Rb is shown in Fig.~9(a). The probe frequency $\nu_P$ was scanned across the transitions $F_g=3\rightarrow F'=2,3$, while the coupling frequency $\nu_C$ was fixed midway between the transitions $F_g=2\rightarrow F'=2,3$. The probe and coupling powers were 0.4~mW and 40~mW, respectively. The beams entered the cell at a small relative angle of about 5~mrad so that, after propagation through the ASC, only the probe beam reached the photodiode and carried the EIT signature.
\begin{figure}
	\centering
	\includegraphics[width=.5\textwidth]{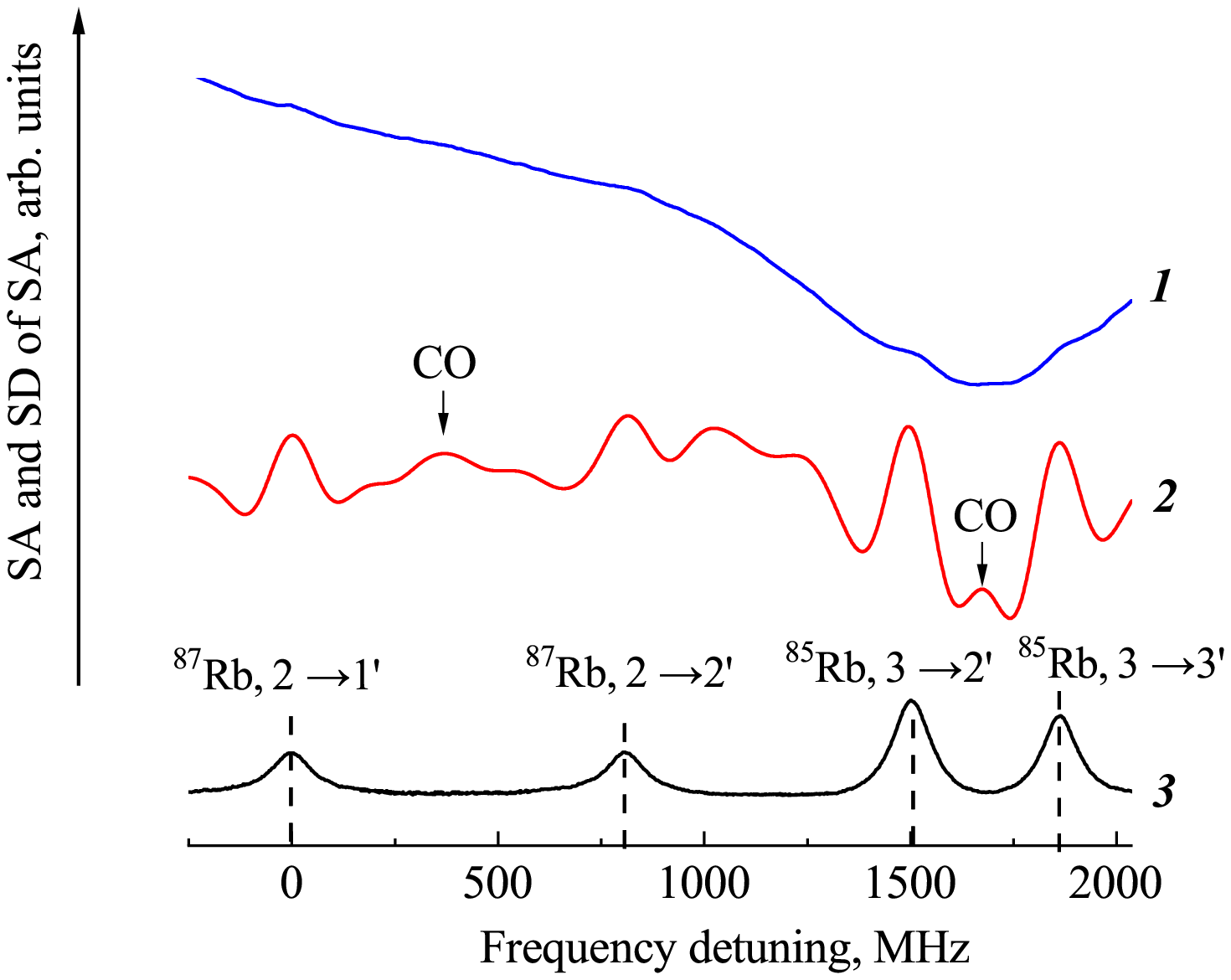}
	\caption{Saturated absorption (SA) spectrum in RRV recorded in a microcell with thickness $L=40\,\mu$m (curve 1). Curve 2 shows the second derivative (SD) of curve 1 (inverted for clarity). Curve 3 presents the reference spectrum. The laser power is $40$\,mW and the cell temperature is $T=190^\circ$C.}
	\label{FIG:8}
\end{figure}

The coupling-field Rabi frequency can be estimated from
\begin{equation}
\frac{\Omega_C}{2\pi}=\gamma_N\sqrt{\frac{I}{8}},    
\end{equation}
where $I$ is the intensity in mW\,cm$^{-2}$ and $\gamma_N/2\pi=5.75$~MHz for the Rb D$_1$ line. This yields $\Omega_C\approx34$~MHz for the present parameters~\cite{29}.

Figure~10 (corresponding to the level scheme of Fig.~9(a)) shows the second derivative of the probe spectrum. Two VSOP resonances are observed on either side of the central EIT feature, corresponding to enhanced absorption at the transitions $F_g=3\rightarrow F'=2,3$. In this configuration, the strong coupling field optically pumps atoms from the ground level $F_g=2$ to $F_g=3$, increasing its population. Between the two VSOP peaks, a narrow EIT resonance appears, characterized by reduced probe absorption and a full width at half maximum of about 12~MHz in the SD representation. These results demonstrate that even in the presence of dense Cs vapor, coherent EIT resonances can be formed in RRV without significant degradation.
\begin{figure}
	\centering
	\includegraphics[width=.3\textwidth]{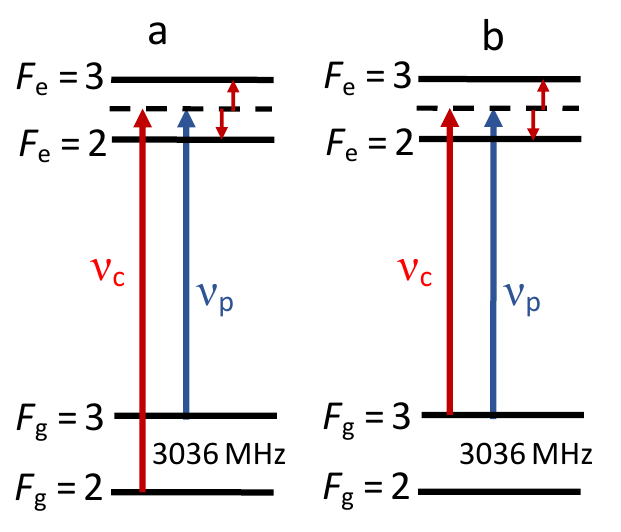}
	\caption{Energy-level scheme of $^{85}$Rb. (a) Coupling and probe lasers with $\lambda \approx 795$~nm form a $\Lambda$-type system. (b) Configuration in which the coupling and probe lasers share the same ground state.}
	\label{FIG:9}
\end{figure}

It is well known that the presence of a buffer gas can increase the interaction time between atoms and the optical field, thereby narrowing the EIT resonance~\cite{25,26}. In the present case, the surrounding Cs vapor effectively plays this role. Figure~10, curve~2 (corresponding to the scheme of Fig.~9(b)), shows a spectrum dominated by optical pumping rather than coherence. Three VSOP resonances with reduced probe absorption are observed at the transitions $F_g=3\rightarrow F'=2,3$ and at the intermediate frequency. Here the coupling radiation transfers population from $F_g=3$ to $F_g=2$, decreasing the population of the probed level. The SD linewidth of these resonances is about 36~MHz, significantly broader than the EIT feature, as expected for a non-coherent process. The EIT linewidth could be further reduced by using phase-coherent probe and coupling beams~\cite{26}.
\begin{figure}
	\centering
	\includegraphics[width=.5\textwidth]{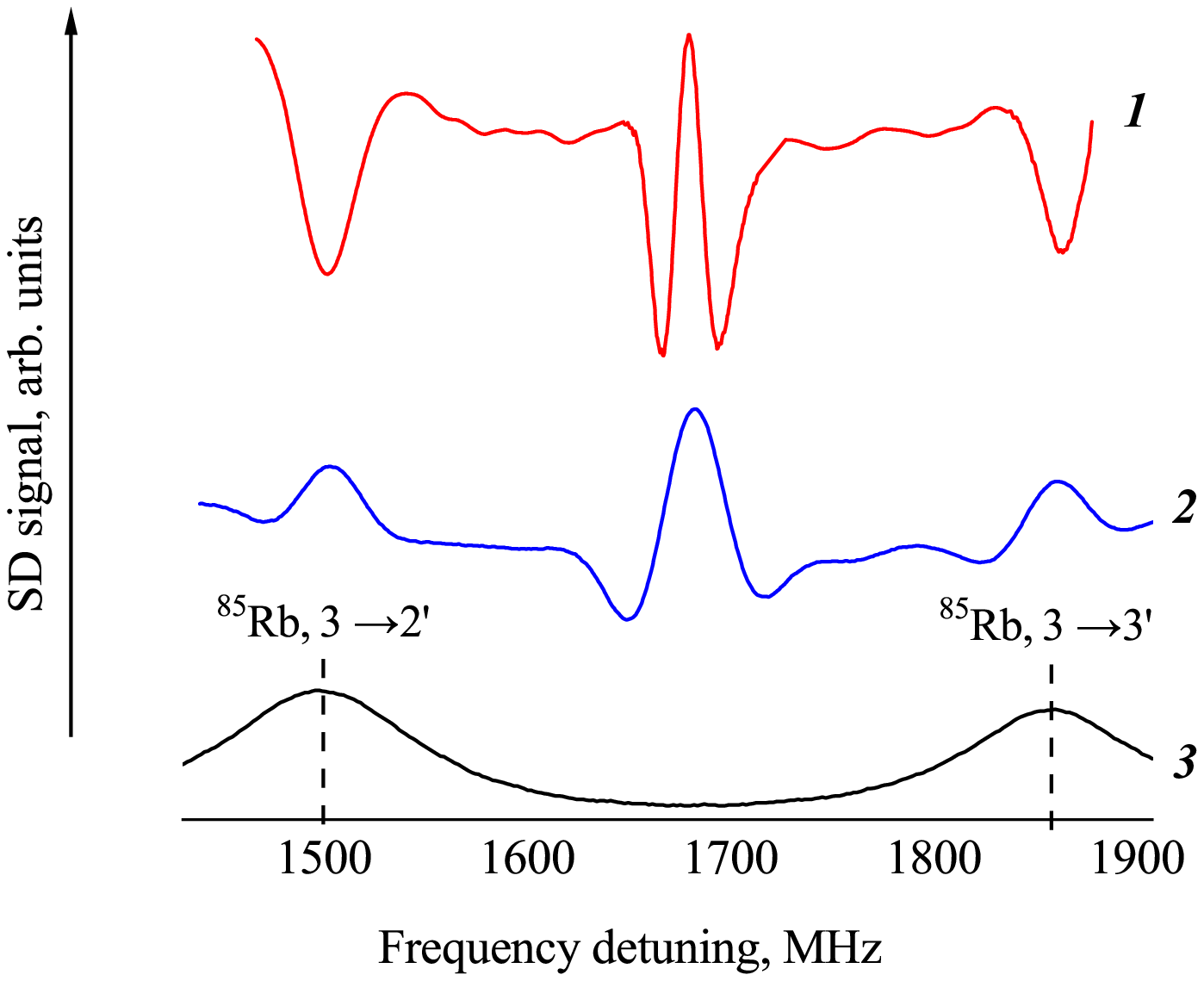}
	\caption{Probe transmission spectra in the presence of the coupling radiation $\nu_{C}$. Curve~1 shows the second-derivative (SD) spectrum containing the EIT resonance and two VSOP resonances. Curve~2 shows the SD spectrum with three VSOP resonances of reduced absorption. Curve~3 presents the reference spectrum.}
	\label{FIG:10}
\end{figure}

The ability to detect spectroscopic signatures from trace atomic species suggests broader applications. For example, saturated-absorption spectroscopy has been used to study vapors containing multiple isotopes, such as Hg~\cite{30}. Using nanocells filled with potassium vapor, Doppler-free spectroscopy has enabled the detection of the rare isotope $^{41}$K at temperatures around $150^{\circ}$C~\cite{31}. Natural potassium also contains only $\sim0.01\%$ of the fermionic isotope $^{40}$K~\cite{32}. In a high-temperature ASC heated to about $440^{\circ}$C, the vapor density of $^{40}$K would reach $\sim10^{13}$~cm$^{-3}$, sufficient for SA measurements of collision-induced broadening. Such systems would allow investigation of fermion--fermion ($^{40}$K--$^{40}$K) and fermion--boson ($^{40}$K--$^{39}$K) collision processes.

\section{Conclusion}

Using an all-sapphire cell (ASC) filled predominantly with Cs vapor ($\sim$99\%), we have demonstrated that high-resolution spectroscopy can be performed on the residual $\sim$1\% fraction of rubidium atoms. Employing ASCs with lengths $L=1$~cm and $L=40~\mu$m, we investigated saturated absorption spectra, electromagnetically induced transparency (EIT) in a $\Lambda$ scheme, and optical pumping processes in the field of two resonant laser beams. 

The dense Cs vapor acts as an effective buffer medium for the residual Rb atoms. Even at a cell temperature of $300^{\circ}$C (corresponding to a Cs vapor pressure of about 2~Torr), velocity-selective optical pumping (VSOP) resonances remain observable in the SA spectra. In contrast, the addition of only 0.11~Torr of argon buffer gas is known to suppress VSOP resonances through velocity-changing collisions~\cite{20}. In the present case, Cs atoms slow the Rb thermal motion and increase the atom--field interaction time, which favors both SA and EIT formation.

These results indicate that trace atomic species inherently present in alkali vapor sources can be exploited for spectroscopic studies without the need for isotopic enrichment. This approach may be particularly useful for systems where the isotope abundance is very low and separation from the dominant isotope is difficult. A notable example is natural potassium vapor, which contains only $\sim0.01\%$ of the fermionic isotope $^{40}$K. In a high-temperature ASC heated to about $440^{\circ}$C, the $^{40}$K density would reach $\sim10^{13}$~cm$^{-3}$, sufficient for SA measurements of VSOP broadening in $^{40}$K--$^{40}$K and $^{40}$K--$^{39}$K collisions, enabling determination of the corresponding collisional cross sections.

Furthermore, related spectroscopic techniques such as microwave--optical double-resonance spectroscopy, recently demonstrated in thin Rb vapor cells~\cite{22}, could also be implemented using residual species in Cs-filled ASCs. Overall, the presence of residual rubidium in cesium vapor cells, often considered an impurity, can instead provide a useful resource for nonlinear and high-resolution spectroscopy, as well as for studies of rare isotopes and collision processes.

These results demonstrate that residual atomic species in alkali vapor cells, often regarded as impurities, can instead serve as a valuable resource for high-resolution and nonlinear spectroscopy, enabling studies of rare isotopes and collisional processes without isotopic enrichment.

\section*{Funding}
The research was supported by the Higher Education and Science Committee of MESCS RA (Research project 1-6/23-I/IPR).

\end{document}